\documentstyle[prl,aps,twocolumn]{revtex}

\begin{document}
\title{Fate of a Bose-Einstein condensate with attractive interaction}
\author{Masahito Ueda$^{1,2}$ and Kerson Huang$^{1.3}$}
\address{$^{1}$ Institute for Theoretical Physics, \\
University of \ California, Santa Barbara, CA 93106-4030 \\
$^{2}$ Department of Physical Electronics, \\
Hiroshima University, Higashi-Hiroshima 739-8257, Japan, and \\
Core Research for Evolutional Science and Technology (CREST), JST\\
$^{3}$ Center for Theoretical Physics and Physics Department,\\
Massachusetts Institute of Technology, Cambridge, MA 02139\\
\rm MIT-CTP \# 2767\qquad cond-mat/9807359}
\date{\today}
\maketitle

\begin{abstract}
We calculate the decay amplitude of a harmonically trapped Bose-Einstein
condensate with attractive interaction via the Feynman path integral. We
find that when the number of particles is less than a critical number, the
condensate decays relatively slowly through quantum tunneling. When the
number exceeds the critical one, a ``black hole'' opens up at the center of
the trap, in which density fluctuations become large due to a negative
pressure, and collisional loss will drain the particles from the trap. As
the black hole is fed by tunneling particles, we have a novel system in
which quantum tunneling serves as a hydrodynamic source.
\end{abstract}

\vspace*{-2pc}  

\noindent {\bf 1. Introduction}

Bose-Einstein condensation has been observed for the alkali atoms $^{87}$Rb~ 
\cite{Anderson}, $^{23}$Na~\cite{Davis}, and $^{7}$Li~\cite{Bradley}, in
magnetic trapping potentials at temperatures well below 10$^{-6}$K. An
important difference among these cases is that the scattering length between
atoms is negative for $^{7}$Li, and positive for the others. This indicates
a predominantly attractive interaction in the former case, and a repulsive
one for the latter. Thus, we expect the $^{7}$Li condensate to be
metastable, and will decay with time. The purpose of this Letter is to
elucidate the mechanism of the decay process.

To be sure, in all three cases the atoms will form metallic crystals in free
space at low temperatures. This means that the short-range interactions are
attractive in all cases, but the trapping potential tends to keep the atoms
apart through zero-point motion, and in the case of positive scattering
length, the atoms do not feel the attractive part of the potential.

We describe the many-boson system with a quantum field $\psi ({\bf r},t)$,
which is governed by the Hamiltonian\cite{Huang} 
\begin{equation}
H=\int d^{3}r\left[ \frac{\hbar ^{2}}{2m}|\nabla \psi |^{2}+\frac{1}{2}%
m\omega ^{2}r^{2}|\psi |^{2}-\frac{u}{2}|\psi |^{4}\right]
\end{equation}
where $m$ is the atomic mass, $\omega $ is the frequency of the external
harmonic potential, and $u=4\pi \hbar ^{2}|a|/m$ characterizes the
interparticle interaction, with negative scattering length $a<0.$ The
transition amplitude from a given initial state $\psi _{0}({\bf r})$ to some
final state $\psi _{1}({\bf r})$ in a time interval $t_{1}$ is given by the
Feynman path integral 
\begin{eqnarray}
T &\equiv &\langle \psi _{1}|\exp \left( -\frac{i}{\hbar }Ht_{1}\right)
|\psi _{0}\rangle  \nonumber \\
&=&\int D\psi D\psi ^{\ast }\exp \left( \frac{i}{\hbar }S[\psi ]\right)
\end{eqnarray}
with boundary conditions $\psi ({\bf r},0)=\psi _{0}({\bf r})$ and $\psi (%
{\bf r},t_{1})=\psi _{1}({\bf r})$. The classical action $S[\psi ]$ is 
\begin{eqnarray}
S[\psi ] &=&\int_{0}^{t_{1}}dt\int d^{3}r\left[ i\hbar \psi ^{\ast }\frac{%
\partial \psi }{\partial t}+\frac{\hbar ^{2}}{2m}\psi ^{\ast }\nabla
^{2}\psi \right.  \nonumber \\
&&\left. -\frac{m\omega ^{2}r^{2}}{2}\psi ^{\ast }\psi +\frac{u}{2}\left(
\psi ^{\ast }\psi \right) ^{2}\right]
\end{eqnarray}
Physically, $\psi _{0}({\bf r})$ and $\psi _{1}({\bf r})$ correspond
respectively to initial and final condensate wave functions, which are
thermal averages of the quantum field. A Feynman path $\psi ({\bf r},t)$
describes a possible history of the condensate wave function.

We calculate the transition amplitude in the semi-classical approximation,
in the formal limit $\hbar \rightarrow 0$. The path integral can be
evaluated in the saddle-point approximation. The saddle-point path, which
corresponds to an extremum of the classical action, obeys the
Gross-Pitaevskii (GP) equation with attractive interaction: 
\begin{equation}
i\hbar \frac{\partial \psi }{\partial t}=-\frac{\hbar ^{2}}{2m}\nabla
^{2}\psi +\frac{m\omega ^{2}r^{2}}{2}\psi -u\psi ^{\ast }\psi ^{2}
\end{equation}
We take as the initial condition $\psi _{0}({\bf r})\!=\!C_{0}\exp$\rlap{$\left(
-r^{2}/2d^{2}\right)$,}\\  where $C_{0}=\sqrt{N/\pi ^{3/2}d^{3}}$. The normalization
is such that $\int d^{3}r\psi _{0}^{2}=N$, where $N$ is the total number of
condensate bosons. The ground state wave function of the harmonic potential has
the same form, but with $d=d_{0}$, where $%
 d_{0}=\left( \hbar /m\omega \right) ^{1/2}$. For our initial state
$d$ is a free parameter.

\medskip
 \noindent {\bf 2. The tunneling path}
\smallskip

The saddle-point path must be chosen to maximize, rather than minimize, the
transition probability. Our Hamiltonian has no lower bound, and the wave
function can grow without limit to attain an ever lower energy. If the
initial wave function is small, however, it must first overcome an energy
barrier presented by the zero-point kinetic term and the external potential.
This it can do via quantum tunneling, which corresponds to a stationary path
in imaginary time \cite{Langer,tHooft,Coleman}. We tailor our initial
discussion to the tunneling process, and put $t=-i\tau .$ Self-consistency
will tell us whether this is the right thing to do. The condensate wave
function in imaginary time is denoted by $\phi ({\bf r},\tau )=\psi ({\bf r,-%
}i\tau ),$ with complex conjugate $\phi ^{\ast }({\bf r},\tau )=\psi ^{\ast
}({\bf r,}i\tau ).$ The final wave function $\phi _{1}$ is determined by the
time evolution. In other words, we follow the path of least action. The
transition probability is given by $|T|^{2}\approx A\exp \left(
-2S_{E}/\hbar \right) $, where $S_{E}$ is the Euclidean action of the
saddle-point path: 
\begin{eqnarray}
S_{E}[\phi ] &=&\int_{0}^{T}d\tau \int d^{3}r\left[ \hbar \phi ^{\ast }\frac{%
\partial \phi }{\partial \tau }-\frac{\hbar ^{2}}{2m}\phi ^{\ast }\nabla
^{2}\phi \right.  \nonumber \\
&&\left. +\frac{m\omega ^{2}r^{2}}{2}\phi ^{\ast }\phi -\frac{u}{2}\left(
\phi ^{\ast }\phi \right) ^{2}\right]
\end{eqnarray}
with $T=it$. We shall not calculate the prefactor $A$ in this paper. Because
our Hamiltonian has a translation invariance with respect to time, the
transition probability should be proportional to $T$ (zero mode) as $%
T\rightarrow \infty $. Thus the transition rate is proportional to $\exp
(-2\hbar ^{-1}S_{E})$ to leading order.

To make all quantities dimensionless, we measure time in units of $\omega
^{-1}$, distance in units of $d_{0}$. Thus, setting $\omega =d_{0}=1$, we
have the imaginary-time GP equation 
\begin{equation}
-2\frac{\partial \phi }{\partial \tau }=-\nabla ^{2}\phi
+r^{2}\phi-g|\phi|^{2}\phi
\end{equation}
where $g=4\pi |a|/d_{0}$, with initial condition $\phi _{0}({\bf r}%
)=C_{0}\exp \left( -\alpha r^{2}/2\right)$, where $\alpha =(d_{0}/d)^{2}$.

Our problem is quite different from the relativistic case \cite
{tHooft,Coleman}, because our equation is first-order in time and not
second-order, and hence it describes diffusion rather than wave motion.
Unlike the relativistic case, the motion here does not obey Newtonian
mechanics in an inverted potential. We therefore do not have either the
instanton \cite{tHooft} or the ``bounce'' \cite{Coleman}.

We now make a modified ``Thomas-Fermi approximation.'' Noting that initially
we have $\nabla ^{2}\phi _{0}=(\alpha ^{2}r^{2}-3\alpha )\phi _{0}$, we make
the replacement $\nabla ^{2}\phi \rightarrow (\alpha ^{2}r^{2}-3\alpha )\phi
.$ Assuming that $\phi $ is real, we can now rewrite the equation in terms
of $\phi ^{2}:$ 
\begin{eqnarray}
\frac{1}{2}\frac{\partial \phi ^{2}}{\partial \tau } &=&-V(r)\phi ^{2}+g\phi
^{4}  \nonumber \\
V(r) &=&\frac{1}{2}\left[ (1-\alpha ^{2})r^{2}+3\alpha \right]
\end{eqnarray}
The solution is given by 
\begin{equation}
\phi ^{2}(r,\tau )=\frac{V(r)}{g\left[ 1+B(r)e^{2\tau V(r)}\right] }
\end{equation}
where $B(r)$ is a monotonically increasing function of $r:$ 
\begin{equation}
B(r)=\frac{V(r)}{g\phi _{0}^{2}(r)}-1
\end{equation}
For large $r$ we have the asymptotic behavior $\phi (r,\tau )\approx \phi
_{0}(r)e^{-\tau V(r)}$ .

\medskip
 \noindent {\bf 3. Critical number of condensate bosons}
\smallskip

If $B(0)>0,$ then $\phi ^{2}(r,\tau )$ will be positive and finite at all
times, showing the self-consistency of the imaginary-time method. The
condition that this be true is $3\alpha /2>gC_{0}^2$, or $N<N_{c},$ with 
\begin{eqnarray}
N_{c}=\frac{3\sqrt{\pi}d}{8|a|}.  \label{critical}
\end{eqnarray}
The total decay rate is $\exp \left( -2S_{E}/\hbar \right) ,$ where 
\begin{equation}
\frac{S_{E}}{\hbar }=\frac{\pi }{g}\int_{0}^{\infty }drr^{2}V(r) \left[ \ln 
\frac{1}{1-f(r)}-f(r)\right]
\end{equation}
where 
\begin{equation}
f(r)=\frac{g\phi _{0}^{2}(r)}{V(r)}=\frac{N}{N_{c}} \frac{\exp\left(-\alpha
r^2\right)}{1+r^2(1-\alpha^2)/(3\alpha)}
\end{equation}
The critical number of condensate bosons $N_{c}$ has been calculated with
various methods \cite{Ruprecht,Baym,Dalfovo}, and all results are of the
same order of magnitude, and not in disagreement with experimental
indications\cite{Hulet1}. Previous estimates of the tunneling probability,
however, give varied physical pictures and mathematical results \cite
{Kagan,Shuryak,Stoof,Ueda}. Our result is closest to that of Ref.~\cite{Ueda}%
.

At each point $r$ in the trap, particles tunnel through an energy barrier
that depends on $r$. Consequently, they ``sweat out'' at different rates at
different points in space. Once over the energy barrier, the saddle-point
path veers onto the real time axis, and the particles evolve according to
the PG equation in real time. We shall comment on their actual fate later.

\medskip
 \noindent {\bf 4. The black hole}
\smallskip

When $N>N_{c}$ \ the initial wave function near the origin is large enough
to go over the energy barrier. This part of the wave function will evolve
according to the PG equation in real time, and we expect that density
fluctuations will become large, with concomitant collisional loss. Thus, a
``black hole'' opens up near the center of the potential, where particles
will be drained at a rapid rate.\ In the imaginary-time formalism, the
existence of the black hole is indicated by the fact that $B(r)$ is negative
at the origin, and increases to zero at $r=b.$ The black hole radius\ $b$ is
determined by the transcendental equation 
\begin{equation}
\phi _{0}^{2}(b)=g^{-1}V(b)
\end{equation}
There is no root unless $N>N_{c}$. Near the critical value we have $%
b^{2}\approx c_{0}(N-N_{c})/N_{c},$ where $c_{0}=3\alpha /(1+2\alpha ^{2})$,
and for large $N$,   $b^{2}$ increases like $\ln N$.

According to our solution, the condensate wave function is independent of $%
\tau $ at the black hole radius, decreases with $\tau $ at any point outside
the radius, and increases with $\tau $ inside. Inside the black hole, the
wave function increases rapidly with decreasing distance, and diverges at $%
r=r_{1}(\tau )$, determined by the condition $1+B\exp (2\tau V)=0$. As $\tau 
$ increases, $r_{1}\rightarrow b$ at an exponential speed, so that $\phi
^{2} $ quickly becomes uniformly divergent just inside the rim of the black
hole. Further into the black hole, $\phi ^{2}$ becomes negative. The last
fact indicates that the imaginary-time formalism is inconsistent inside the
black hole, and we should go to a real-time description.

\medskip
 \noindent {\bf 5. Negative pressure and instability}
\smallskip

Reverting to real time inside the black hole, we write $\psi ({\bf r},t)=%
\sqrt{n}e^{i\theta }$, and obtain a pair of hydrodynamic-like equations\cite
{Stringari_R}. One of these is the continuity equation $\partial n/\partial
t+\nabla \cdot \left( n\nabla \theta \right) =0,$ and the other reads $%
\partial \theta /\partial t+\Omega =0$, where 
\begin{equation}
\Omega =\frac{1}{2}(\nabla \theta )^{2}+\frac{1}{2}\left( r^{2}-\frac{1}{%
\sqrt{n}}\nabla ^{2}\sqrt{n}\right) -gn
\end{equation}
Using the modified Thomas-Fermi approximation, we replace the second term on
the right side by $V(r)$. The steady-state solution $\partial \theta
/\partial t=0$ gives $\Omega =0$, or $\frac{1}{2}v_{s}^{2}+V-gn=0,$ where $%
{\bf v}_{s}=\nabla \theta $ is the superfluid velocity. This is like
Bernoulli's law in hydrodynamics, but with negative pressure $P=-gn$.
Solving for the superfluid velocity we obtain 
\begin{equation}
v_{s}=\sqrt{2\left[ gn(r)-V(r)\right] }
\end{equation}
The region in which this is real is precisely the black hole. Outside this
region $v_{s}$ becomes pure imaginary, and we are back in the tunneling
regime.

Since the pressure is negative, however, the system must be highly unstable
with respect to density fluctuations, which were ignored in our modified
Thomas-Fermi approximation. The instability must lead to nonuniform collapse
into pockets of extremely high density, with superfluid turbulence. We must
also take into account the fact that inelastic collisions, such as
three-body molecular recombinations, become increasingly likely with
increasing density, and these collisions will expel particles from the trap 
\cite{Kagan,Dodd}. This region is therefore a cauldron of density
fluctuations and collisional loss, and truly a black hole.

\medskip
 \noindent {\bf 6. Outlook}
\smallskip

The trapped Bose-Einstein condensate with attractive interaction offers many
exciting possibilities for theoretical and experimental study.

(1) The opening and closing of a black-hole region, whose size depends on
the number of particles, furnishes a mechanism for the phenomenon of
condensate oscillation suggested recently \cite{Hulet2,Kagan2}. A
quantitative study will necessitate a dynamical treatment.

(2) In a kinetic theory, we expect that the particles ``sweating'' out
throughout the trap, due to tunneling, will create a steady flow into the
black hole, and drain out eventually due to collisions. Here, for the first
time, we have a system in which macroscopic tunneling serves as hydrodynamic
source. The kinetic theory in this case will have to bridge quantum
tunneling and fluid mechanics.

(3) By plugging the black hole with a laser beam of variable size, one can
create a ``faucet'' that controls the draining of the condensate. The
remaining condensate, which decays slowly via tunneling, should support a
larger maximum number.

\medskip

\centerline{\hfill\hbox to 2cm{\hrulefill}\hfill}   
\smallskip

We thank Randy Hulet for sharing with us some of his data before
publication, the organizers of the Workshop on Bose-Einstein condensation at
the Institute for Theoretical Physics at Santa Barbara, 1998, for making
this work possible, and the participants of the Workshop for discussions.
One of us (KH) was supported in part by DOE cooperative agreement
DE-FC02-94ER40818.

\end{document}